\documentclass{ws-mpla}
\usepackage{url,cite,amsfonts}

\begin{document}

\title{A simple variational principle \\
for classical spinning particle \\
with  anomalous magnetic momentum}
\author{S.A.Pol'shin}

\address{Institute for Theoretical Physics \\
NSC Kharkov Institute of Physics and Technology \\
Akademicheskaia St. 1, 61108 Kharkov, Ukraine \\
polshin.s at gmail.com}

\maketitle

\begin{abstract}
We obtain Bargmann-Michel-Telegdi equations of motion of classical spinning particle  using Lagrangian variational principle with Grassmann  variables.

\end{abstract}

\ccode{PACS 2008 Nos.: 75.10.Hk 11.30.Pb 11.30.Ef }

The famous Bargmann-Michel-Telegdi (BMT) equations~\cite{BMT} describe the motion of classical spinning particle with anomalous magnetic moment in the constant and homogeneous electromagnetic field (see also~\cite{Jack} for textbook treatment). However, their validity in the inhomogeneous field is unclear and the corresponding variational principle is unknown. To resolve these difficulties, in the present note we use appropriate generalization of a method proposed by Ravndal~\cite{Rav} to  describe the motion of classical spinning particle with no anomalous magnetic moment using four-dimensional Grassmann variables. Unlike the usual approach (see~\cite{GGit} and references therein), the Ravndal's one avoids any use of fifth coordinate which considerably simplifies the theory. Later the motion in Schwarzschild~\cite{Riet}, Taub-NUT~\cite{Baleanu98,Banu}, Coulomb~\cite{Kozoriz}, and in the combined gravimagnetic~\cite{Mohseni} backgrounds as well as in the torsion field~\cite{Rumpf}  were considered within the Ravndal's approach.

Working in the same approach, in the present letter we formulate an appropriate variational principle for classical spinning particle  with anomalous magnetic momentum and show that the corresponding equations of motion coincide with BMT ones at the lowest orders of Grassmann analytic expansion. See also~\cite{GitSaa-mpla,GitSaa} for the description of anomalous magnetic moment within the usual five-dimensional approach.

Let $\mathbb{R}_{S}$ be the supercommutative superalgebra with infinite number of generators and  $B=C^\infty (\mathbb{R})\otimes \mathbb{R}_{S}$, then $B=B_0 \oplus B_1$, where $B_0$ and $B_1$  are even and odd parts of $B$ respectively (see~\cite{Rogers} for basic definitions of superspaces and superalgebras). Let $x^\mu \in B_0$ and $\xi^\mu \in B_1$ for each $\mu=0,\ldots,3$, so we deal with superspace $B^{4,4}\ni (x^\mu, \xi^\mu)$, where $x^\mu$ are coordinates of configuration space and $\xi^\mu$ are Grassmann variables which describe the spin degrees of freedom.

Let $F^{\mu\nu}(x)$ be a "super-electromagnetic field", i.e. for each $\mu,\nu$ it is a function $B^{4,0}\rightarrow B$ which obeys the second pair of "super-Maxwell equations"
\begin{equation}\label{smaxw}
  \varepsilon^{\mu\nu\rho\sigma} \partial^E_\nu F_{\rho\sigma}=0,
\end{equation}
where $\partial^E_\mu$ is the so-called even derivative~\cite{Rogers}.

Consider the Lagrangian (cf.~\cite{Rav} for the case $\mu=e$)
 \begin{equation}\label{eq1}
L=\left(\frac{m}{2}\dot{x}\vphantom{x}^\mu \dot{x}_\mu
-\frac{1}{4}\xi^\mu \dot{\xi}_\mu\right)+eA_\mu \dot{x}\vphantom{x}^\mu+
\frac{\mu'}{2m}F^{\mu\nu}S_{\mu\nu},
\end{equation}
where $\mu'/2m$ is the magnetic moment of a particle, $S_{\mu\nu}=\frac{1}{2}\xi_\mu \xi_\nu$ is the spin tensor and overdots mean the derivatives wrt proper time $s$ of the particle. Impose the constraint (cf.~\cite{Rav})
\begin{equation}\label{eq2}
\xi_\mu \dot{x}\vphantom{x}^\mu=0
\end{equation}
and let $\lambda$ be the corresponding Lagrange multiplier. Observe that the chain rule holds for superdifferentiable functions (Theorem 4.4.2 of~\cite{Rogers}). Then  varying the action $\int (L+\lambda \xi_\mu \dot{x}\vphantom{x}^\mu)\, ds$ wrt $x$ and $\xi$ we obtain equations of motion (cf.~\cite{Rav} for the case $\mu'=e$, then $\lambda=0$)
\begin{eqnarray}
m\ddot{x}\vphantom{x}^\mu=eF^{\mu\nu}\dot{x}_\nu +\frac{\mu'}{2m}\eta^{\mu\kappa}
(\partial^E_\kappa F^{\rho\sigma}) S_{\rho\sigma} -(\partial^E_\nu \lambda)\xi^\mu
\dot{x}\vphantom{x}^\nu \label{eq3} \\
\dot{\xi}\vphantom{\xi}^\mu =\frac{\mu'}{m} F^{\mu\nu}\xi_\nu-2\lambda
\dot{x}\vphantom{x}^\mu. \label{eq4}
\end{eqnarray}
The above equations should preserve the constraint~(\ref{eq2}), so using~(\ref{smaxw}) we obtain
\begin{equation}\label{eq5}
\lambda \dot{x}\vphantom{x}^\mu\dot{x}_\mu=\frac{\mu'-e}{2m}F^{\mu\nu} \dot{x}_\mu \xi_\nu.
\end{equation}

For $f\in B^{4,4}$ let $\underline{f}$ be the nonzero term of lowest degree wrt the generators of $\mathbb{R}_{S}$. Then  $\underline{x}$ is of 0th degree, $\underline{\xi}$ is of 1th degree, $\underline{F}$ is of 0th degree and it is easily seen that it obeys the second pair of usual Maxwell equations since $\underline{\partial^E_\mu F_{\nu\rho}}=\partial_\mu \underline{F}\vphantom{F}_{\; \nu\rho}$, so it may be interpreted as usual electromagnetic field. Suppose $\underline{x}^\mu \underline{\dot{x}} \vphantom{x}_{\: \mu}=1$, then using~(\ref{eq3}),(\ref{eq4}) we obtain the closed system of equations of motion
\begin{eqnarray}\label{eq7}
m\underline{\ddot{x}}\vphantom{x}^\mu
=e\underline{F}^{\mu\nu}\underline{\dot{x}}\vphantom{x}_\nu \\
m\underline{\dot{S}}\vphantom{S}^{\mu\nu}=\mu' \underline{F}\vphantom{F}^{\rho[\nu}
\underline{S}\vphantom{S}_{\: \rho}^{\: \ \mu]} +(\mu'-e)
\underline{F}\vphantom{F}^{\rho\sigma} \underline{\dot{x}}\vphantom{x}_{\: \rho}  \underline{S}\vphantom{S}_{\: \sigma}^{\: \ [\mu} \underline{\dot{x}}\vphantom{x}^{\nu]} \label{eq8}
\end{eqnarray}
which are equivalent to BMT ones.

\section*{Acknowledgment}

I am grateful to Yu.P.Stepanovsky for helpful discussions and constant interest on my work.

\end{document}